\newcommand{\be}{\begin{equation}}\newcommand{\ee}{\end{equation}}
\newcommand{\bea}{\begin{eqnarray}}\newcommand{\eea}{\end{eqnarray}}

\newcommand{\p}[1]{(\ref{#1})}

\newcommand{\bD}{\overline D}

\newcommand{\cU}{{\cal U}}

\newcommand{\cF}{{\cal F}}
\newcommand{\cZ}{{\cal Z}}
\newcommand{\bcZ}{{\overline{\cal Z}}}

\newcommand{\bz}{{\bar z}}
\newcommand{\bxi}{{\bar\xi}}
\newcommand{\bpsi}{{\bar\psi}}

\documentclass[12pt]{article}
\usepackage{amscd,amsmath,amssymb}

\begin{document}

\thispagestyle{empty}
\vspace{2cm}
\begin{flushright}
\end{flushright}
\begin{center}
{\Large\bf Generic N=4 supersymmetric hyper-K\"ahler sigma models in D=1}
\end{center}
\vspace{1cm}

\begin{center}
{\large\bf S.~Bellucci${}^{a}$, S.~Krivonos${}^{b}$, A.~Shcherbakov${}^{a,b}$ }
\end{center}

\begin{center}
${}^a$ {\it INFN-Laboratori Nazionali di Frascati,
Via E. Fermi 40, 00044 Frascati, Italy}

\vspace{0.2cm}

${}^b$ {\it Bogoliubov  Laboratory of Theoretical Physics, JINR, 141980 Dubna,
Russia}

\vspace{1cm}
bellucci@lnf.infn.it, krivonos@theor.jinr.ru, ashcherb@lnf.infn.it

\end{center}
\vspace{2cm}

\begin{abstract}
We analyse the geometry of four-dimensional bosonic manifolds arising within the context of $N=4, D=1$ supersymmetry.
We demonstrate that both cases of general hyper-K\"ahler manifolds, i.e.
those with translation or rotational isometries, may be supersymmetrized in the
same way. We start from a generic N=4 supersymmetric three-dimensional action and perform
dualization of the coupling constant, initially present in the action. As a result,
we end up with explicit component actions for $N=4, D=1$ nonlinear sigma-models with hyper-K\"ahler geometry
(with both types of isometries) in the target space. In the case of hyper-K\"ahler geometry with
translational isometry we find that the action possesses an additional hidden $N=4$ supersymmetry, and therefore
it is $N=8$ supersymmetric one.

\end{abstract}

\newpage
\setcounter{page}{1}
\section*{Introduction}
It has been known for a long time that the target space geometry of the sigma model is intimately related with the number
of supersymmetries it possesses. In particular, Bruno Zumino showed that $(2,2)$ supersymmetry in $D=1$ requires the bosonic
part of a Lagrangian to be a K\"ahler manifold \cite{1}. Later on, Alvarez-Gaume and Freedman \cite{2} proved
that $(4,4)$ supersymmetry further restricts the target space geometry to be hyper-K\"ahler (HK). Next, the analysis of
the supersymmetric sigma-models with Wess-Zumino terms \cite{hkt} and heterotic
$(4,0)$ supersymmetric sigma models \cite{{3},{4}} brought about hyper-K\"ahler geometries with torsion (HKT)\footnote{
The same type of bosonic target HKT geometry as in the $D=1$ case was present in the $N=8, D=1$ analytic bi-harmonic superspace, see e.g. \cite{bis}
and references therein.}.
Apart from
their evident application to non-linear sigma-models, HK and HKT geometries arise also in the moduli spaces for a certain
class of black holes \cite{4}, in the target space of a bound state of a D-string and D-five-branes, etc. Unfortunately, all
these applications, though very interesting, are rather complicated. Moreover, the mathematical description
of supersymmetric sigma models with HK and/or HKT target space geometries is quite involved. Therefore it seems to be a promising
idea to simplify everything in such a way as to provide the simplest theory where the HK geometry arises as a consequence of
supersymmetry, and where all main properties of the theory can understood. Clearly enough, supersymmetric
mechanics should be a good choice, in this respect.

The supersymmetric mechanics with $N=4$ supersymmetry possesses a number of specific features which make it selected
with respect, not only to its higher-dimensional counterparts, but also to mechanics with a
different number of supersymmetries. Firstly,
$N=4, D=1$ supersymmetry is rather simple. Moreover, just in the $N=4, D=1$ case the most general action may be easily
written in terms of superfields as an integral over the whole superspace (in close analogy with $(2,2)$ supersymmetry in $d=2$).
Secondly, all known $N=4$ supermultiplets in $D=1$ are off-shell, so the corresponding actions can be written
in standard superspace (see e.g. \cite{ikl} and refs. therein). One should stress that just in $N=4, D=1$ superspace
one may define a new class of nonlinear supermultiplets which contain a functional freedom in the defining
relations \cite{{ks},{di},{bk41}}. Let us recall that we formulated the problem of
how to describe $N = 4$ and $N = 8$, $D = 1$ sigma models with HK metrics in the target space in \cite{lectures}, advocating the use
of nonlinear supermultiplets.
Finally, in one dimension there is a nice duality between cyclic variables in Lagrangian
and coupling constants. Indeed, if some one dimensional Lagrangian has a cyclic variable, say $\phi$, then the corresponding
conserved momentum $p_\phi$ acquires a constant value $m$. Performing a Routh transformation over $\phi$ we will get a theory
with a smaller number of bosonic fields but with a coupling constant $m$. Obviously, this procedure may be reversed to dualize
the coupling constant $m$ into a new bosonic field $\phi$. Clearly enough, the resulting Lagrangian will possess an isometry
with the Killing vector $\partial/\partial \phi$. In what follows we will heavily use just these features of supersymmetric mechanics.

It is known that four dimensional bosonic hyper-K\"ahler manifolds with (at least) one isometry may be divided into two types
that are in fact distinct from each other\footnote{Here we closely  follow \cite{basf}.}.
The first kind, which is sometimes called translational (or triholomorphic), corresponds to a Killing
vector with self-dual covariant derivatives. In the supersymmetric case the translational isometry commutes with supersymmetry.
For the four dimensional
hyper-K\"ahler manifolds with a translational isometry there is a preferred coordinate system where the bosonic sigma model action
reads \cite{hk1}
\be\label{hk1}
S_1 = \int dt \left[ \frac{1}{g} \left( \dot\phi +\omega_i \dot{x}^i \right)^2 + g \left( \eta_{ij} {\dot x}^i \dot{x}^j \right)\right],
\ee
where $\eta_{ij}$ is a flat three dimensional metric and $\omega_i(x^j)$ and $g(x^i)$ are constrained to satisfy the conditions
\be\label{hk1a}
\partial_i g = \pm \epsilon_{ijk}\partial_j \omega_k.
\ee
It immediately follows from \p{hk1a} that the metric $g(x^i)$ satisfies the three dimensional Laplace equation. Let us observe that one may
always choose the ``gauge'' $\omega_3=0$ by a proper redefinition of the field $\phi$.

The second type of four dimensional hyper-K\"ahler manifolds, which are called manifolds with rotational isometry,
encompasses all other Killing vector fields. Once again, one may find a preferred coordinate system in which the sigma model
action gets the simplest form \cite{hk2}
\be\label{hk2}
S_2=\int dt\left[ \frac{1}{\Psi_u} \left( \dot\phi +i  \Psi_z \dot{z} - i  \Psi_{\bar z} \dot{\bz}\right)^2+
{\Psi_u} \left( {\mbox e}^\Psi \dot{z} \dot{\bar z} + {\dot u}^2\right)\right].
\ee
Here, additionally, the function  $\Psi=\Psi(z,{\bar z},u)$ satisfies the Toda equation
\be\label{hk2a}
 \Psi_{z  {\bar z}} + \left( {\mbox e}^{\Psi}\right)_{uu} =0.
\ee

The purpose of the present Letter is to construct $N=4$ supersymmetric extensions of both bosonic HK metrics \p{hk1} and \p{hk2}.
The line of our construction looks as follows. It is rather easy to realize that after removing the cyclic variable $\phi$ in
both actions \p{hk1} and \p{hk2}, we will get the three dimensional mechanics with a conformally flat metric in the case of \p{hk1}
and with a more complicated metric in the case of \p{hk2}. Therefore we start from the  general $N=4, D=1$ three dimensional mechanics,
properly constrain it to get  the needed three dimensional bosonic manifolds, and then perform a dualization of a coupling constant,
initially present in the action, to reproduce the full action.
To close this Section let us mention that the $N=4$ supersymmetric mechanics for the action with translational isometry \p{hk1}
has been already constructed: on the component level in \cite{{G},{ks},{bks11}} and in the harmonic superspace in \cite{di}.
As for the action with rotational isometry  \p{hk2}, to the best
of our knowledge, no explicit supersymmetric action has been constructed. In the next Sections we are going to consider both cases on the
same footing. Surprisingly, for the case \p{hk1} the system we found admits $N=8, D=1$ supersymmetry. In the following, we simplify
our presentation by considering only the bosonic parts of the corresponding actions. The complete expressions, including all fermionic
terms, may be easily restored, if needed.

\section*{From three dimensional to hyper-K\"ahler sigma models}
The most general three dimensional sigma model with $N=4, D=1$ supersymmetry can easily be constructed within
standard $N=4, D=1$ superspace which may be parameterized by the following coordinates:
$ t, \theta_i,{\bar\theta}{}^i; i=1,2$.
We choose as our basic superfields the ordinary $N=4$ chiral superfield $\cZ$ obeying the conditions
\be\label{sf1}
D^i \bcZ=0, \quad \bD_i \cZ=0,
\ee
and the so called ``old tensor" supermultiplet  \cite{leva} which may be described by a real superfield $\cU$ subjected to
the following constraints:
\be\label{sf2}
D^i D_i \;\cU = \bD^i\bD_i \;\cU=0,
\ee
where the $N=4$ spinor covariant derivatives $D^i,\bD_i$ obey the standard super-Poincar\'e algebra
$$\left\{ D^i, \bD_j \right\} =2i \delta^i_j \partial_t.$$ The chiral superfield $\cZ$ describes two physical bosons $z,\bar z$,
four fermions $\psi^i, \bpsi_i$ and two auxiliary bosonic fields $A, \bar A$ which may be defined as\footnote{As usual, $|$ denotes
the restriction to $\theta^i=\bar\theta_i=0$.}
\be\label{comp1}
z=\cZ|,\; {\bar z}=\bcZ|,\; \psi^i =D^i\cZ|, \; \bpsi_i=\bD_i \bcZ|, \; A= D^iD_i \cZ|,\; {\bar A}=\bD_i \bD^i \bcZ|.
\ee
Concerning the ``old tensor" supermultiplet, it comprises one physical boson $u$, once again four fermions $\xi^i,\bxi_j$ and a triplet of
auxiliary components $A^{(ij)}$
\be\label{comp2}
u=\cU|,\; \xi^i =D^i \cU|, \; \bxi_i =\bD_i \cU|, \; A_{(ij)} =i \left[ D_{(i},\bD_{j)}\right] \cU|.
\ee
What is extremely important for our construction is that among the components of the superfield $\cU$ there is a constant $m$ \cite{leva}.
Indeed, from the basic constraints \p{comp2} it immediately follows that
\be\label{g}
\frac{\partial}{\partial t} \left[ D^i, \bD_i\right] \cU =0 \; \Rightarrow \; \left[ D^i, \bD_i\right] \cU=4m=\mbox{ const}.
\ee
The most general sigma model action may be easily written in the full $N=4, d=1$ superspace as
\be\label{a1}
S= -\int dt d^4 \theta \; \cF(\cZ, \bcZ, \cU)\equiv -\frac{1}{4}\int dt D^2 \bD{}^2 \; \cF(\cZ, \bcZ, \cU),
\ee
where $\cF(\cZ, \bcZ, \cU)$ is an arbitrary real function of $\cZ, \bcZ, \cU$. After passing to the components \p{comp1},\p{comp2}
and eliminating the auxiliary fields by their equations of motion, the bosonic part of the action \p{a1} takes the following form:
\be\label{a2}
S_{bos}=\int dt \left[\left( F_{uu} {\dot u}^2 - 4F_{z\bz}{\dot z}{\dot\bz}\right)  -F_{uu}m^2 +2im\left(F_{uz}{\dot z} -
F_{u\bz}\dot\bz \right)  \right].
\ee
Now we have at hands the most general three dimensional sigma model action. The next task is to put a proper restriction on the
prepotential $\cF(\cZ, \bcZ, \cU)$ to reproduce the three dimensional part of the metrics \p{hk1} and \p{hk2} and to perform the dualization
of the coupling constant $m$.
\subsection*{HK sigma model with translational isometry}
For the HK sigma model with translational isometry the bosonic three dimensional part of the action should be conformally flat
as in \p{hk1}. It is immediately clear from \p{a2} that conformal flatness is achieved if
\be\label{eq1}
F_{z\bz}=-F_{uu} \; \Rightarrow F_{uu}+F_{z\bz}=0.
\ee
Thus, the necessary conditions to reproduce \p{hk1} is to choose the prepotential $\cF$ to be a three dimensional harmonic function.
In this case, the three dimensional metric reads
\be\label{eq1a}
g=F_{uu}
\ee
which, as consequence of \p{eq1},  obeys  the three dimensional Laplace equation. Thus, we partially achieved the needed action.
All that we still need is to get the full four dimensional action \p{hk1}. Fortunately, the action \p{a2} already contains the
coupling constant $m$ which may be dualized into a fourth bosonic field. Let us supply the action \p{a2} with an additional term
\be\label{a2a}
{\tilde S}_{bos} = S_{bos} + \int dt\;m \dot\phi.
\ee
Varying \p{a2a} over the new bosonic field $\phi$ we will simply recover that $m=const$. But if we will instead vary the action \p{a2a} over
$m$, which is now an independent variable, we will immediately get
\be\label{m1}
m = \frac{1}{2F_{uu}}  \left[\dot\phi+2i\left(F_{uz}{\dot z} - F_{u\bz}\dot\bz \right)\right].
\ee
Plugging \p{m1} back into the action \p{a2a} we will finally get
\be\label{finA}
{\tilde S}_{bos} =\int dt\left\{ F_{uu} \left(  {\dot u}^2 + 4 {\dot z}{\dot\bz}\right) +
\frac{1}{4F_{uu}}  \left[\dot\phi+ 2i\left(F_{uz}{\dot z} - F_{u\bz}\dot\bz \right)\right]^2 \right\}.
\ee
Comparing the action \p{finA} with \p{hk1} one may find that they completely coincide (modulo unessential numerical factors)
after passing in the action \p{hk1} to the complex coordinates $z=x^1+ix^2$, choosing the gauge $\omega_3=0$ and
with conditions \p{hk1a} being  solved exactly.

Thus, we constructed an $N=4$ supersymmetric mechanics which possesses HK geometry with translation isometry in its bosonic target space.
Surprisingly, the condition we have to impose on the prepotential \p{eq1} to be a harmonic function is completely sufficient
to realize an additional $N=4$ supersymmetry  which commutes with the manifest one and still preserves the action \p{a1} \cite{bikl}.
Therefore, we conclude that our action \p{finA}, with all fermionic terms being restored, provide us with $N=8,D=1$ supersymmetric
mechanics with HK geometry in the bosonic sector.
\subsection*{HK sigma model with rotational isometry}
Now we will turn to the second type of HK metrics \p{hk2}. Once more, comparing the kinetic term of the action \p{a2} with the
corresponding part of the action \p{hk2} one may conclude that now, in order to reproduce the action \p{hk2}, we have to impose
the following constraints on the prepotential $F$:
\be\label{eq2}
F_{z\bz}=-\mbox{e}^{\Psi} F_{uu}, \; F_{uu} = \partial_u \Psi .
\ee
One may check that the integrability condition of the system \p{eq2} results in the equation
\be\label{eq2a}
\frac{\partial}{\partial u} \left[\Psi_{ z  {\bar z}} + \left( {\mbox e}^{\Psi}\right)_{uu}\right] =0,
\ee
which is just a weak variant of the condition \p{hk2a}, necessary in order to get the HK metric. As the last step we will perform
the same dualization of the constant $m$ as in the previous section. As a result we end up with the following bosonic action:
\be\label{finB}
{\hat S}_{bos} =\int dt\left\{ \Psi_{u} \left(  {\dot u}^2 + 4 \mbox{e}^\Psi {\dot z}{\dot\bz}\right) +
\frac{1}{4\Psi_{u}}  \left[\dot\phi+ 2i\left(\Psi_{z}{\dot z} - \Psi_{\bz}\dot\bz \right)\right]^2 \right\},
\ee
where we partially integrated \p{eq2} as $F_u =\Psi$.
Clearly, we have the same action as in \p{hk2}. Hence, we conclude that the action \p{finB}, with all
fermionic terms being restored, correctly reproduces $N=4,D=1$ supersymmetric
mechanics possessing the HK geometry with rotational isometry in the bosonic sector.

\section*{Discussions and Outlook}
The study of supersymmetric quantum mechanics models endowed with $N = 4,8$ supersymmetry represents one of the most
up-to-date and prolific direction of development. While building upon the researches initiated in \cite{leva},
the new developments aim at a more complete understanding of the
structure of the corresponding higher-dimensional supersymmetric field theories, as well as of
the AdS(2)/CFT(1) correspondence.

In this paper we constructed $N=4, D=1$ supersymmetric sigma models with HK geometry in the bosonic target
space which possess (at least) one translational/rotational isometry. In the case of HK geometry with
translational isometry we found that the action possesses an additional hidden $N=4$ supersymmetry and therefore
it is $N=8$ supersymmetric \cite{bikl}.
 We also explicitly demonstrated that the conditions which select these types of
HK metrics follow from the invariance under $N=4$ supersymmetry.

Being almost manifestly supersymmetric, our construction leaves one serious question unanswered. Indeed, we
obtained the fourth bosonic physical field through the dualization of the coupling constant, which from the beginning
is present in one of the supermultiplets we started from. Thus, the question is: how is $N=4$ supersymmetry realized
on this new field $\phi$? The transformations properties of $\dot\phi$ under $N=4$ supersymmetry may be immediately
found from its definition \p{m1}. We have explicitly checked that on-shell the time derivative could be taken off
and the transformation law of $\phi$ is local under $N=4$ supersymmetry
\be
\delta\phi=8i \left( \epsilon^i D_i F_u -\bar\epsilon_i \bD{}^i F_u \right).
\ee
So, at least we deal with the local realization of $N=4$ supersymmetry, but it is still unclear, whether it might be realized off-shell.
Another interesting  question concerns the explicit realization of the hidden $N=4$ supersymmetry in case of translational isometry
and the possibility of existence of hidden $N=4$ supersymmetry in the case of the action with rotational isometry.
Of course, in this respect the superfield off-shell formulation could be very useful. Unfortunately, at present we do not have
a such formulation at hands.

Finally, we would like to stress that considered $D=1$ sigma-models are much simpler then their $D=4$
and even $D=2$ counterparts. Moreover, they are completely ready for quantization. In principle, one may hope
to find the exact spectrum for some peculiar solutions of three-dimensional Laplace and Toda equations.
Let us observe that even the three dimensional action \p{a2} is interesting, owing to the very specific interaction terms.
This is especially intriguing in view of the existence of two examples of integrable systems (with translational isometry)
where interactions are defined in such a way \cite{{gib},{ners}}. Hopefully, the extended supersymmetry will not spoil
integrability. We hope to report the corresponding results elsewhere.

\section*{Acknowledgements}

The authors are grateful to F.~Delduc and E.~Ivanov for  useful correspondences.
S.K. would like to thank the INFN--Laboratori Nazionali di Frascati for the
warm hospitality extended to him during the course of this work.

This work was partially supported by the European Community's Marie Curie Research Training Network under contract
MRTN-CT-2004-005104 Forces Universe, by INTAS under contract 05-7928 and by grants RFBR-06-02-16684, DFG~436 Rus~113/669/03.

\end{document}